\documentclass[twocolumn,aps,prx,floatfix,longbibliography,groupedaddress]{revtex4-2}
\usepackage{amsmath,bbm}
\usepackage{graphicx}
\usepackage{multirow}
\usepackage{hyperref}
\usepackage{amsfonts}
\usepackage{amsbsy}
\usepackage{xcolor}
\usepackage{soul}
\usepackage{graphicx}
\usepackage{mathtools}
\usepackage{bm}

\renewcommand{\vec}[1]{\mathbf{#1}}
\newcommand{\vecg}[1]{\bm{#1}}

\newcommand{\Pc}{\mathcal{P}}

\newcommand{\bra}[1]{\ensuremath{\left\langle #1 \right\vert}}
\newcommand{\ket}[1]{\ensuremath{\left\vert #1 \right\rangle}}
\newcommand{\unitvec}[1]{\hat{\mathbf{{#1}}}}
\renewcommand{\vec}[1]{\mathbf{#1}}
\newcommand{\id}{\mathbb{I}}
\newcommand{\half}{\hbox{$1\over2$}}

\begin{document}
\title{Quantum single-photon control, storage, and entanglement generation\\ with planar atomic arrays }

\date{\today}
\author{K.~E.~Ballantine}
\email{k.ballantine@lancaster.ac.uk}
\author{J.~Ruostekoski}

\email{j.ruostekoski@lancaster.ac.uk}

\affiliation{Department of Physics, Lancaster University, Lancaster, LA1 4YB, United Kingdom}

\begin{abstract}

While artificially fabricated patterned metasurfaces are providing paradigm-shifting optical components for classical light manipulation,  strongly interacting,  controllable, and deterministic quantum interfaces between light and matter in free space remain an outstanding challenge.
Here we theoretically demonstrate how to achieve quantum control of both the electric and magnetic components of an incident single-photon pulse by engineering the collective response of a two-dimensional atomic array. High-fidelity absorption and storage in a long-lived subradiant state, and its subsequent retrieval, are achieved by controlling classically or quantum mechanically the ac Stark shifts of the atomic levels and suppressing the scattering during the absorption. 
Quantum wavefront control of the transmitted photon with nearly zero reflection is prepared by coupling the collective state of the array to another photon in a cavity and by engineering a Huygens' surface of atoms using only a single coherent standing wave.
The proposed schemes allow for the generation of entanglement between the cavity, the lattice, and hence the state of the stored, reflected or transmitted light, and for quantum-state transfer between the cavity and propagating photons. 
Bipartite entanglement generation is explicitly calculated between a stored single-photon excitation of the array and the cavity photon.
We illustrate the control by manipulating the phase, phase superposition, polarization, and direction of a transmitted or reflected photon, providing quantum-optical switches and functional quantum interfaces between light and atoms that could form links in
a larger quantum information platform.
\end{abstract}

\maketitle

\section{Introduction}

Modular quantum networks require versatile quantum systems to act as individual nodes, as well as coherent links to transfer information between them~\cite{ChoiEtAlNature2008,Duan10,Ritter12}.
Light-atom interfaces utilizing
two-dimensional (2D) atomic arrays exhibit strong cooperative scattering, provide rich control of collective excitations~\cite{Jenkins2012a,Bettles2016,Facchinetti16,Yoo2016,Shahmoon,Bettles2017,Perczel2017a,Plankensteiner2017,Asenjo-Garcia2017a,Mkhitaryan18,Orioli19,Williamson2020b,Cidrim20,
Parmee2020,Ballantine21PT}, and are emerging as key many-body quantum elements which could fill this role.
Highly collimated light emission allows for the arrays to be efficiently linked in free space~\cite{Facchinetti18,Grankin18,Javanainen19} and even to form single-photon quantum antennas~\cite{Grankin18,Ballantine20ant}. This eliminates a major loss channel of spontaneous emission in undesired directions that limits the efficiency of single-photon schemes employing single atoms or disordered atom clouds in quantum information processing~\cite{HAM10}. The effective 1D propagation also leads to greatly enhanced optical cross-sections, which are not achievable for single atoms in free space, while coupling the single atom, e.g., to light propagation in 1D waveguides is plagued with photon losses.  
Similarly, an efficient light-matter interface is difficult to obtain using, e.g., the DLCZ  protocol of spin waves in free-space disordered ensembles~\cite{HAM10}, requiring very high optical thickness -- a condition  that can be exponentially improved in cooperatively coupled atomic arrays~\cite{Asenjo-Garcia2017a}.
Moreover, atomic planar arrays can exhibit subradiant states, with lifetimes many orders of magnitude longer than in a single atom, that can be selectively targeted~\cite{Facchinetti16}, utilized in sensing~\cite{Facchinetti18}, metrology~\cite{Kramer2016,Qu19}, or in preparing Bell states~\cite{Guimond2019}.
Further information processing, such as two-photon quantum gates, could potentially be engineered using Rydberg-state nonlinearities~\cite{Zhang21,Cardoner21,Bekenstein2020}.
Experimental realizations of atom arrays are now achievable with increasing positional control~\cite{Weitenberg11,Glicenstein2020,Dordevic21,Xia15,Lester15,Kim16,Cooper18}, with the subradiant spectral narrowing of transmitted light below the fundamental single-atom linewidth having recently been observed~\cite{Rui2020}. 

The revolutionary potential of thin 2D surfaces in controlling and manipulating the flow of classical light has been recognized with artificial fabricated metasurfaces~\cite{Decker15,Yu14,Luo18}, where versatile functionalities
are obtained via microscopic patterned coating and not by modifying the geometric shape of the material. However, the quest for reaching the quantum regime in such systems is still an outstanding challenge~\cite{Solntsev21}. Artificial materials also suffer from many disadvantages compared to natural atoms, including absorptive heating losses and fabrication inconsistencies. Cooperatively interacting planar atomic arrays can be engineered to exhibit optical properties previously only attained to artificial metamaterials and not natural media, such as strong magnetic interactions at optical frequencies~\cite{Ballantine20Huygens,Alaee20}. Moreover, an atomic surface consisting of superpositions of induced synthetic electric and magnetic dipoles can generate 
a Huygens' surface~\cite{Ballantine20Huygens,Ballantine21}, representing a physical implementation of Huygens' principle of light propagation~\cite{Huygens,Love1901}.

\begin{figure*}[t]
  \centering
   \includegraphics[width=2\columnwidth]{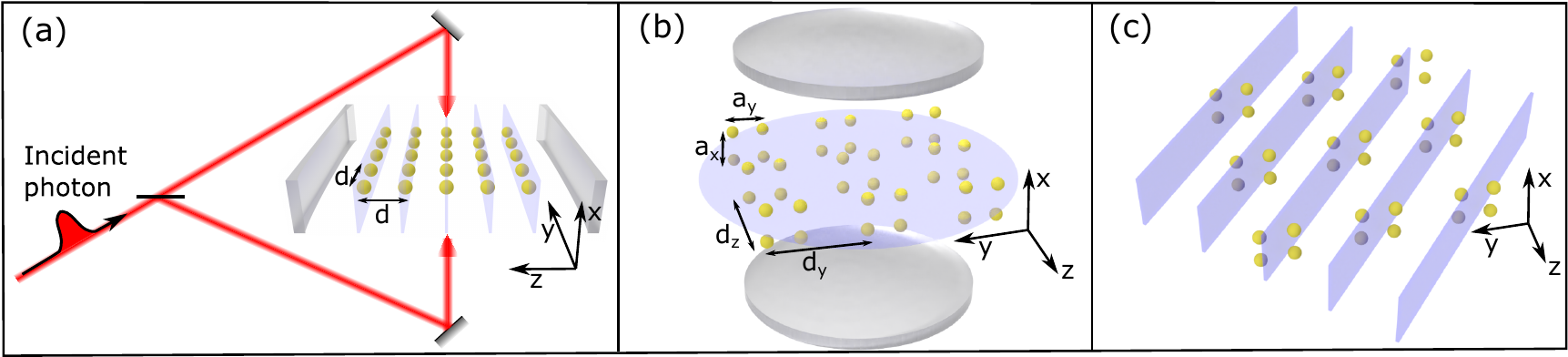}
  \caption{Example geometries of atomic lattice and cavity. In all cases, level shifts can be engineered with quantized or classical standing waves inducing ac Stark shifts to control (a) the storage, (b) reflection, or (c) transmission properties of an incident single photon along the $x$ axis. (a) A regular square lattice with lattice constant $d$. High-fidelity absorption and storage of a single-photon pulse is controlled by a spatially uniform shifting of the $m=\pm 1$ levels, which could be achieved with a quantized cavity field or classical standing wave, e.g., propagating in the lattice plane as shown. (b) A bilayer lattice consisting of a rectangular array of square unit cells centered in the $yz$ plane. A cavity photon can be used to engineer a spatially uniform distribution of level shifts,  changing the resonant mode to control the phase of reflected light, or controlling an atomic Huygens' surface to switch between different wavefront-shaping configurations. (c)  A simple scheme to engineer the atomic Huygens' surface, controlled by the cavity photon in (b), relying only on ac Stark shifts induced by a single classical standing wave. }
  \label{fig:geo}
\end{figure*}

Here we expand the classical optics functionalities of ultrathin, flat 2D metasurfaces to the quantum regime by employing planar atomic arrays. Instead of relying on the manipulation of single-particle degrees of freedom, we show how the \emph{collective} radiative excitations of atoms can be engineered to achieve single-photon quantum control of both its \emph{electric and magnetic} components. Examples include the deterministic absorption and storage of a single-photon pulse as a subradiant excitation and the manipulation of the phase, polarization, and direction of reflected or transmitted photon.  We demonstrate how this engineering can be achieved via an interacting qubit, such as a cavity photon, generating entanglement and enabling quantum-state transfer between the cavity and outgoing photons.
This allows for the storage, extreme wavefront control, and coherent transfer of single photons as part of a distributed quantum network. 
 
We first show how a time-dependent single-photon pulse can be fully absorbed, stored in, and deterministically released from, a subradiant state in a 2D atomic lattice. 
Previous proposals have shown how excitations in 2D lattices can be almost completely transferred into a strongly subradiant state under continuous steady-state illumination by controlling the orientations of the atomic dipoles by level 
shifts~\cite{Facchinetti16,Facchinetti18} or by optimizing driving~\cite{Manzoni2017}. Experimentally, substantial subradiant mode steady-state excitation has been observed in planar dipolar arrays~\cite{Jenkins17,Rui2020}, while in random atomic ensembles small numbers of photons have exhibited very long radiative lifetimes~\cite{Guerin_subr16,Ferioli21}. Such techniques, however, are not directly applicable for a specific protocol of high-fidelity storage of a photon pulse where the incoming photon is absorbed by first driving the transition dipole moment that necessarily radiates, resulting in significant losses.

In our protocol, ac Stark shifts~\cite{gerbier_pra_2006} of the atomic levels create a linear Zeeman splitting
during the pulse excitation, causing the collective atomic polarization to rotate until it points perpendicular to the plane, a subradiant state where scattering is strongly suppressed. Full absorption is achieved by first splitting the incident pulse at a beamsplitter and illuminating the lattice symmetrically from both sides. The pulse duration is chosen such that the incident and scattered light destructively interfere to suppresses any outgoing light. 
The excitation can be stored with high efficiency  by turning off the level shifts when the subradiant state is maximally occupied, and later retrieved by turning them back on. Furthermore, a quantized cavity photon can be used to induce the required shifts, generating entanglement between the cavity and the lattice excitation. We demonstrate this by calculating bipartite entanglement generation between a stored single-photon excitation of the array and the cavity photon in Appendix~\ref{sec:entanglement}.

Extreme wavefront control of a transmitted photon with a quantum-coherent, entanglement-preserving protocol can be achieved with an interacting qubit tuning the resonance of an atomic Huygens' surface. This takes the ideas of metasurfaces to the realm of quantum nanophotonics by demonstrating a birefringent variable waveplate, photon steering, and quantum-optical switch of the phase of a reflected photon from the atoms that simultaneously act as an ultrathin flat electric and magnetic mirror.
The quantum state of the auxiliary cavity photon can be used to switch between two different wavefront-shaping configurations, generating entanglement between the cavity photon and the phase, polarization, and direction of the transmitted light, and allowing the quantum state to be transferred from the cavity to the propagating photon. 
We propose a simple scheme to realize an atomic Huygens' surface -- and therefore, a phased-array antenna with in-principle arbitrary wavefront manipulation of the transmitted light -- where the required level shifts are prepared by only a single standing-wave laser.
Our scheme allows for the single-photon quantum control as part of a larger quantum architecture, with the state of the cavity depending on prior operations, and the photons stored, sorted, and redirected into outgoing modes in a way that preserves entanglement.

\section{Atom light interactions}
\label{sec:ints}

We consider a regular array of atoms, with one atom per site, and a $J=0\rightarrow J^{\prime}=1$ transition. 
Two different lattice geometries are explored, shown in Fig.~\ref{fig:geo}, a simple square lattice and a bilayer lattice, both in the $yz$ plane, while an incident single-photon probe field propagates in the $x$ direction. Since the $z$ quantization axis of the  $J^\prime =1$, $m=0,\pm 1$ levels is chosen to be orthogonal to the photon propagation direction, the $z$ polarization component of the incident field drives the transition to the $m=0$ level, while the $y$ component drives a combination of the $m=\pm 1$ transitions.

The reduced many-body density matrix for atoms, describing the evolution of a pure single-photon excitation, can be decomposed as $\rho=\ket{\Psi}\bra{\Psi}+p_G\ket{G}\bra{G}$ with $p_G$ the probability that the excitation has decayed to the state $\ket{G}=\prod_j\ket{g_j}$ with all atoms in their respective electronic ground state $\ket{g_j}$, and we ignore the effects of recoil~\cite{Robicheaux19}. The single excitation sector is described by the state~\cite{Ballantine20ant} $\ket{\Psi}=\sum_{\mu j} \Pc_{\mu}^{(j)}\hat{\sigma}_{\mu j}^{+}\ket{G}$ with $\hat{\sigma}_{\mu j}^{+}=\ket{e_{\mu j}}\bra{g_j}$ the raising operator to level $m=\mu$ on atom $j$. We write the $3N$ excited-state amplitudes, corresponding to three polarization components labeled by $\mu$ on $N$ atoms labeled by $j$, in terms of a single vector $b_{3j+\mu-1}=\Pc_\mu^{(j)}$, which evolves according to $\dot{\vec{b}}=i(\mathcal{H}+\mathcal{H}^{\prime})\vec{b}+ \vecg{\zeta}(t)$, where we approximate the driving by the incident field by a coherent single-photon pulse $\zeta_{3j+\mu-1}=\unitvec{e}_\mu^\ast \cdot \unitvec{e}_{\rm in} f(\vec{r}_j) \exp{(-t^2/\tau^2)}$,  with spatial amplitude $f(\vec{r})$ and polarization $\unitvec{e}_{\rm in}$. The non-Hermitian coupling matrix~\cite{Lee16} $\mathcal{H}$ has diagonal entries $\Delta\Omega+i\gamma$, where $\gamma=\mathcal{D}^2k^3/(6\pi\hbar\epsilon_0)$ is the single-atom linewidth, $\mathcal{D}$ is the reduced dipole matrix element, and $\Delta\Omega=\Omega-\omega$ is the experimentally controlled detuning of the incident photon frequency $\Omega=ck$ from the single-atom resonance $\omega$. The off-diagonal elements are given by $\mathcal{H}_{3j+\mu-1,3k+\nu-1}=i\xi \unitvec{e}_\mu^{\ast}\cdot \mathsf{G}(\vec{r}_j-\vec{r}_k)\unitvec{e}_\nu$, where $\xi=6\pi\gamma/k^3$ and $\mathsf{G}$ is the standard dipole radiation kernel such that the field at a point $\vec{r}$ from a dipole $\vec{d}$ at the origin is $\epsilon_0\vec{E}(\vec{r})=\mathsf{G}(\vec{r})\vec{d}$~\cite{Jackson}. 
Additional engineered level shifts $\Delta_{\mu}^{(j)}=\omega-\omega^{(j)}_{\mu}$ for each level $\mu$ of atom $j$ are incorporated in the diagonal matrix $\mathcal{H}^{\prime}$.

The collective response of the array can be engineered by using a control field, consisting of either a classical coherent laser or a quantized cavity mode, to induce ac Stark shifts of the atomic levels~\cite{gerbier_pra_2006}. We take this field to be a standing wave $\hat{\vec{E}}_s(\vec{r})=\mathcal{E}_0\hat{a}_s\cos{(\vec{k}_s\cdot\vec{r}+\phi)}\unitvec{e}_s$ with $\mathcal{E}_0=\sqrt{\hbar\omega_s/(2V\epsilon_0)}$, wavevector $\vec{k}_s$, polarization $\unitvec{e}_s$, frequency $\omega_s$, phase $\phi$, and mode volume $V$. In the engineering of the Huygens' surface, and in the simplest case of introducing level shifts to control single-photon storage, we use coherent standing-wave laser fields $ \mathcal{E}_0\hat{a}_s\rightarrow  \mathcal{E}_s$ with amplitude $\mathcal{E}_s$, while for coherent quantum control of single-photon storage, reflection phase, and switching between Huygens' surface configurations, the lattice is placed in a cavity with photon annihilation operator $\hat{a}_s$. The control field weakly couples the $J^\prime=1$ levels to other high-lying states, leading to an effective dressing and shifting their resonance frequencies. The resulting interaction has the form $\hat{H}=  \sum_{\mu j} U(\vec{r}_j,\omega_s,\mu,\unitvec{e}_s)\mathcal{E}_0^2 \hat{a}^{\dagger}_s \hat{a}_s\hat{\sigma}^{ee}_{\mu j}$, with $\hat{\sigma}^{ee}_{\mu j}=\ket{e_{\mu j}}\bra{e_{\mu j}}$, where again for the classical beam we take $\mathcal{E}_0^2\hat{a}_s^\dagger\hat{a}_s\rightarrow |\mathcal{E}_s|^2$. Here the function $U$ incorporates an overall dependence on the spatial variation of the control field, as well as both the detuning of the frequency $\omega_s$, and the overlap of the polarization $\unitvec{e}_s$ with the transition from each level $\mu$  (see Appendix~\ref{sec:Stark})~\cite{Schmidt16,Rosenbusch09,LeKien13}. For the case of quantum control, the cavity supports a single quantized mode, initialized in a superposition of zero or one photon states, i.e.\ $\sum_{n=0,1}c_n\ket{n}$ with $\hat{a}_s^\dagger \hat{a}_s\ket{n}=n\ket{n}$.

To understand the response of the lattice and the effect of the control field, we describe the dynamics in terms of the collective excitation eigenmodes that are given by the eigenvectors $\vec{v}_n$ of $\mathcal{H}$~\cite{Sokolov2011,Jenkins2012a,Castin13,Jenkins_long16,Lee16}. The corresponding eigenvalues $\delta_n+i\upsilon_n$ provide the collective line shifts $\delta_n$  and the collective linewidths $\upsilon_n$ with $\upsilon_n<\gamma$ defining subradiant modes that, for sufficiently large arrays, can have lifetimes many orders of magnitude longer than a single atom. In the examples that we consider, the response is well described by the interaction of only two modes. One is the collective uniform in-phase mode $\Pc_1$, with all dipoles oscillating in the $y$ direction, in the plane of the lattice, which couples strongly to normally incident, $y$-polarized, plane-wave or Gaussian illumination. 
The other mode $\Pc_2$, which may or may not couple directly to the incident field, differs according to the particular application we consider. 
The modes $\Pc_{1,2}$ of $\mathcal{H}$ are not eigenmodes of the full evolution matrix $\mathcal{H}+\mathcal{H}^{\prime}$ in the presence of non-uniform level shifts $\Delta_{\mu}^{(j)}$, and for an appropriate choice of these shifts the two modes can be coupled together (see Appendix~\ref{sec:tmm}). 
This coupling is illustrated by the two-mode model~\cite{Facchinetti16,Facchinetti18}  
\begin{equation}\label{eq:tmm}
\begin{pmatrix}
\dot{\Pc_1} \\ \dot{\Pc_2}
\end{pmatrix} = i\left[(\mathcal{H}_{\rm tm}+\Delta\Omega\id) +\mathcal{H}_{\rm tm}^{\prime}\right]\begin{pmatrix}
\Pc_1 \\ \Pc_2
\end{pmatrix}+\begin{pmatrix}
\zeta_1(t) \\ \zeta_2(t)
\end{pmatrix},
\end{equation}
where $\id$ is the $2\times 2$ identity matrix and
\setlength{\arraycolsep}{1pt}
\begin{equation}
\mathcal{H}_{\rm tm}=\begin{pmatrix}
\delta_{1} + i \upsilon_1 && 0 \\ 0 && \delta_{2} + i \upsilon_2 
\end{pmatrix}, \,
\mathcal{H}_{\rm tm}^{\prime}=\begin{pmatrix}
\tilde{\delta} && -i\bar{\delta} \\ i\bar{\delta} && \tilde{\delta} 
\end{pmatrix}.
\end{equation}
The resonances are shifted by $\tilde{\delta}$, resulting from a uniform shift of all the levels, while the contribution of the variation in the atomic level shifts is encapsulated in the effective coupling $\bar{\delta}$. These parameters can be controlled by engineering the level shifts across the lattice via either the cavity photon or a classical field. 

\section{Single-photon storage}
\label{sec:storage}

We show how a pulsed single photon can be efficiently absorbed and stored in a delocalized subradiant collective excitation by considering a single-layer square lattice with spacing $d$. 
Unlike in the case of 1D chains of atoms~\cite{Williamson2020,Holzinger21}, in 2D arrays strongly subradiant modes cannot be directly driven by incident fields. 
The process can be qualitatively modeled by Eq.~(\ref{eq:tmm}) when we take $\Pc_2$ to be the uniform in-phase oscillation of all dipoles in the $x$ direction, perpendicular to the lattice plane~\cite{Facchinetti16}. This mode is extremely subradiant because, in the limit of infinite atom number, a sub-wavelength lattice can only scatter perpendicular to the plane in the zeroth-order Bragg peak. When the dipoles point out of the plane, however, they do not radiate in this direction, and so no scattering is possible~\cite{CAIT,Javanainen19}. While the excitation can eventually decay from a finite array, light on average is scattered many times towards the edges of the lattice before it can escape, leading to a dramatic narrowing of the linewidth. However, the mode is in general difficult to excite as incident light cannot match the uniform phase and $x$ polarization. In the two-mode model of Eq.~(\ref{eq:tmm}), these conditions correspond to $\zeta_2=0$, $\upsilon_2\ll \gamma$ (with $\lim_{N\rightarrow\infty}\upsilon_2=0$).

To efficiently absorb and store the photon, we take a Gaussian, $y$-polarized, input pulse such that the incident light strongly drives the in-plane $\Pc_1$. 
However, in the absorption process light first induces the dipole moment on the atoms and these dipole moments unavoidably start radiating, leading to scattering and potential losses. 
When the lattice is illuminated from one direction, the collimated forward scattered light from the 2D array overlaps spatially with the incident field, leading to destructive interference and reducing transmission, but the losses due to back-scattering can still 
be significant. In order to suppress radiation from the array in both directions while absorption is ongoing, the pulse is passed through a beam splitter to symmetrically illuminate the lattice from both sides~\cite{Roger15}, as shown in Fig.~\ref{fig:geo}(a). The pulse duration is chosen so that the scattered field has approximately equal magnitude to the incident field, to maximize destructive interference. If $\bar{\delta}=0$ then, as the incident field falls away, the scattering continues, but the interference no longer occurs, and light is quickly reradiated from $\Pc_1$, with the lattice returning to its ground state. However, an ac Stark shift~\cite{gerbier_pra_2006} can lift the degeneracy of the  $m=\pm 1$ levels, causing an effective linear Zeeman splitting and coupling $\Pc_y$ to $\Pc_x$ at each atom. If the level splitting is uniform on all atoms, the coherent phase of the dipoles is maintained, and the uniform in-plane mode is coupled to the similarly uniform-phase out-of-plane mode $\Pc_2$, with coupling $\bar{\delta}=(\Delta_{+}^{(j)}-\Delta_{-}^{(j)})/2$ depending on the strength of the Zeeman splitting, and $\tilde{\delta}=(\Delta_{+}^{(j)}+\Delta_{-}^{(j)})/2$ (see Appendix~\ref{sec:tmm})~\cite{Facchinetti16,Facchinetti18}. The excitation can then be stored in this subradiant mode by turning off the level shifts, removing the coupling between $\Pc_2$ and $\Pc_1$. 

\begin{figure}[htbp]
  \centering
   \includegraphics[width=\columnwidth]{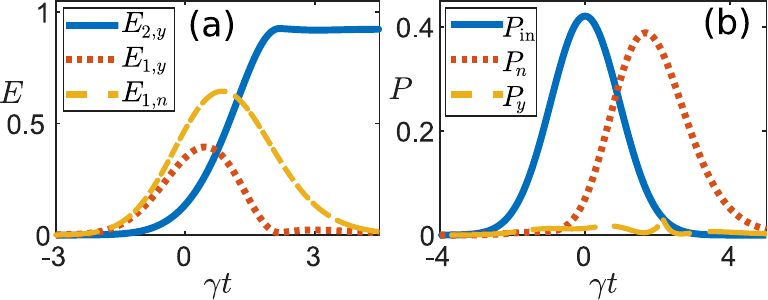}
  \caption{Storing a single-photon pulse in a subradiant state. (a)
Energy $E_i$ contained in each mode $i$, in units of $\hbar\Omega$, in the case that level shifts $\bar{\delta}=0.65\gamma$ are present (y) or not (n). In the former case shifts are turned off at $\gamma t=2.3$. (b) Incident power ($P_{\rm in}$) of single-photon pulse, in units of $\gamma\hbar\Omega$, and outgoing power in both directions in case where shifts are present ($P_y$) or absent ($P_n$). Numeric simulations for $31\times 31$ lattice with $d=0.54\lambda$ and $\gamma\tau=1.9$. }
  \label{fig:storage}
\end{figure}

We first assume that the uniform level shifts for $m=\pm1$ for each atom are induced by a classical coherent control field. As an example we consider $\vec{k}_s=(\pi/d)\unitvec{z}$ and circular polarization $\unitvec{e}_s=\unitvec{e}_{+}$. This field has equal intensity on all atoms but has opposite polarization projection onto the $m=\pm 1$ states, causing the desired splitting~\cite{Schmidt16,Rosenbusch09,LeKien13}. The resulting mode energy and transmitted power are shown in Fig.~\ref{fig:storage} for the case where level shifts are present or absent. At a time $\gamma t=2.3$, when the $\Pc_2$ mode is maximally occupied, the level splittings are turned off, and the excitation is stored successfully. Numerically, we find $93\%$ of the incident pulse is absorbed. While previous proposals~\cite{Facchinetti16,Facchinetti18} and experiments~\cite{Jenkins17,Rui2020}  have shown how to drive steady-state subradiant modes of 2D dipolar arrays under continuous illumination, the combination here of symmetric illumination and the synchronization of the level shifts with the time-dependent incident pulse allows for the total deterministic absorption of a whole single-photon pulse.

\begin{figure}[htbp]
  \centering
   \includegraphics[width=\columnwidth]{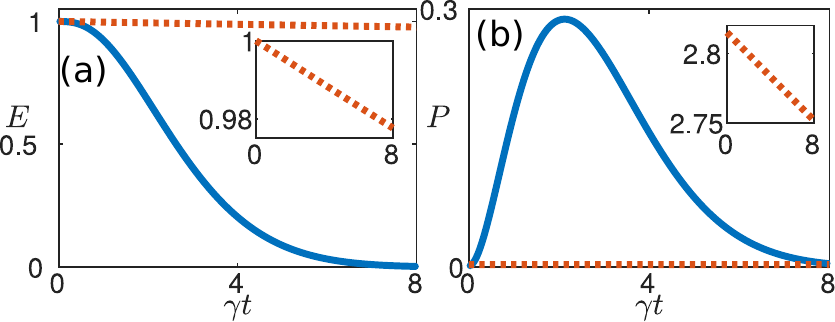}
  \caption{Release of a single-photon excitation initially stored in the subradiant state. (a) Energy remaining in collective excitation, in units of $\hbar\Omega$, for the uniform level shifts generating $\bar{\delta}=0.5\gamma$ and the out-coupling of the photon (solid line), as well as in the absence of the shifts $\bar{\delta}=0$ (dashed and inset). (b) Output power of single-photon pulse, in units of $\gamma\hbar\Omega$, for $\bar{\delta}=0.5\gamma$ (solid) and $\bar{\delta}=0$ (dashed and inset in units $\gamma\hbar\Omega\times 10^{-3}$). Numerical simulation of $31\times 31$ atom array with $d=0.54\lambda$.   }
  \label{fig:decay}
\end{figure}

The stored single-photon excitation in the subradiant state $\Pc_2$ can subsequently be retrieved in a similar scheme by turning the level shifts on and coupling $\Pc_2$ to $\Pc_1$. Figure~\ref{fig:decay} shows (a) the remaining excitation energy a time $t$ and (b) the emitted pulse profile, starting from an initial energy $\hbar\Omega$. The release of the single-photon excitation can be thought of as the time-reversal of the absorption of the incident pulse. However, the emitted pulse is not exactly Gaussian, leading to a discrepancy between absorption and the time-reversed release and resulting in the non-perfect $93\%$ efficiency of storage. Tailoring the shape of the incident photon pulse to more closely match the retrieved pulse, therefore, provides a simple means to optimize and further improve the storage and retrieval fidelity. Numerical integration of Eq.~\eqref{eq:tmm}, taking $\zeta_2$ to have the same time-dependence as the emitted single-photon pulse, shows an efficiency $\approx 99.95\%$.  A release of subradiant excitations by applying spatially-dependent level shifts (causing inhomogeneous broadening, instead of rotation of dipoles considered here) was experimentally achieved in Ref.~\cite{Ferioli21}.

While absorption, storage, and release of the single-photon pulse can most simply be achieved by using a classical field of a laser or microwave, we can extend the scheme to quantum control by using the interaction with a single qubit to induce the effective Zeeman splitting. This leads to entanglement generation between the cavity and the array, and allows the storage and retrieval to form part of a larger quantum-information processing architecture. Here, we take this qubit to be a quantized cavity photon, with $\vec{k}_s=(\pi/d)\unitvec{z}$ and $\unitvec{e}_s=\unitvec{e}_{+}$ as previously in this section, as shown in Fig.~\ref{fig:geo}(a). The initial state is a separable state consisting of an incident single-photon pulse, $\vecg{\zeta}(t)$, and the cavity superposition state, $c_0\ket{0}+c_1\ket{1}$.
The subspaces of the combined system when the cavity is projected onto the state with zero or one photon then evolve very differently, with the single-photon quickly being reradiated in the absence of level shifts, or coupled into $\Pc_2$ and stored successfully when the cavity is occupied. The state of the cavity and the incident pulse become strongly entangled (see Appendix~\ref{sec:entanglement}). While in the previous case the classical field controlling the level shift was switched off to store the photon, this is not possible when a cavity photon generates the level shift. Instead, a classical field with equal and opposite level shift can now be switched on, leading to net-zero coupling. In the alternate case that the cavity is empty, this classical field will lead to a non-zero coupling $-\bar{\delta}$. However, while this will cause some temporary occupation of $\Pc_2$, the two modes will remain coupled, and the excitation will not be stored. The cavity photon can also be used to produce the level shifts necessary to release a stored photon, resulting in entanglement between the cavity mode and the single-photon pulse that is either stored or retrieved.

\section{Quantum-coherent control of magnetic response}

\subsection{Engineering optical magnetism}
\label{sec:magnetism}

So far, we have considered the excitation of uniform electric dipole oscillations. Atomic arrays can also be engineered to have higher multipole resonances, including strong collective magnetic dipole and electric quadrupole responses~\cite{Ballantine20Huygens,Alaee20}. We now turn to the engineering and control of these modes in a bilayer lattice, consisting of square unit cells with atoms spaced at $x=\pm a_x/2$, $y=nd_y\pm a_y/2$, and $z=md_z$, for integer $n$, $m$, as shown in Fig.~\ref{fig:geo}(b). Here we also take $d_y/a_y$ to be an integer such that the non-Bravais lattice could be realized by selectively filling sites of a regular rectangular bilayer lattice~\cite{Koepsell20,Gall21}, but the proposed geometry could also be achieved by using a double-well superlattice in the $y$ direction~\cite{Kangara18}, or with optical tweezers~\cite{Barredo16}. Tightly spaced unit cells could be achieved, for example, by using a transition analogous to $^3P_0$ to $^3D_1$ in bosonic $^{88}$Sr, with a wavelength $\lambda=2.6\,\mu\mathrm{m}$, allowing for deeply sub-wavelength atomic positioning~\cite{Olmos13}.

\begin{figure}[htbp]
  \centering
   \includegraphics[width=\columnwidth]{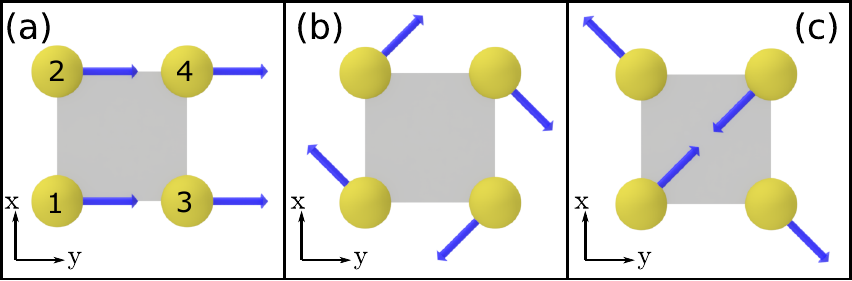}
  \caption{Selected excitation modes of a square unit cell. The arrows point to the direction of the excited electric dipoles of the $J=0\rightarrow J'=1$ transition in each atom. (a) Electric dipole mode with combined electric dipole pointing in $y$ direction. (b) Magnetic dipole mode and (c) electric quadrupole mode with multipole moment pointing in the $z$ direction in both cases. Labels show the order of level shift parameters give in Fig.~\ref{fig:polhs}. }
  \label{fig:modes}
\end{figure}

To analyze the collective response, we first consider the scattering from an individual square unit cell consisting of the four atoms around a fixed $(nd_y,md_z)$. The scattered light from this sub-wavelength structure can be decomposed into electric and magnetic multipole contributions~\cite{Jackson},
\begin{equation}
\vec{E}_s = \sum_{l=0}^\infty\sum_{m=-l}^{l}\left(\alpha_{\mathrm{E},lm}\vec{\Psi}_{lm}+\alpha_{\mathrm{B},lm}\vec{\Phi}_{lm}\right),
\label{eq:harmonics} 
\end{equation}
for vector spherical harmonics $\Psi_{lm}$, $\Phi_{lm}$, with $\alpha_{\mathrm{E},lm}$ giving the strength of the electric dipole ($l=1$), quadrupole ($l=2$) etc.\ and $\alpha_{\mathrm{B},lm}$ the corresponding magnetic multipoles. The formation of three of these multipole modes in terms of the electric dipole moment of each individual atom within a unit cell is shown in Fig.~\ref{fig:modes}. In the first, uniform in-phase $y$ polarization on each atom leads to a pure electric dipole, when all four individual dipoles interfere constructively. Another mode, shown in Fig.~\ref{fig:modes}(b), has a net-zero electric dipole moment, but instead forms a magnetic dipole, with polarizations pointing in the azimuthal direction around the unit cell approximating a current loop~\cite{Ballantine20Huygens,Ballantine20Toroidal,Alaee18}, while Fig.~\ref{fig:modes}(c) shows one of several possible electric quadrupole modes. Strong light-mediated interactions between unit cells lead to collective modes of the entire lattice, consisting of a uniform repetition of a single multipole on each unit cell. In the following discussions, we refer to such collective modes, with the repeating similar pattern at each unit cell, as the uniform electric dipole, magnetic dipole, and electric quadrupole modes, respectively. 

\subsection{Phase-cat state}
\label{sec:cat}

\begin{figure}[htbp]
  \centering
   \includegraphics[width=\columnwidth]{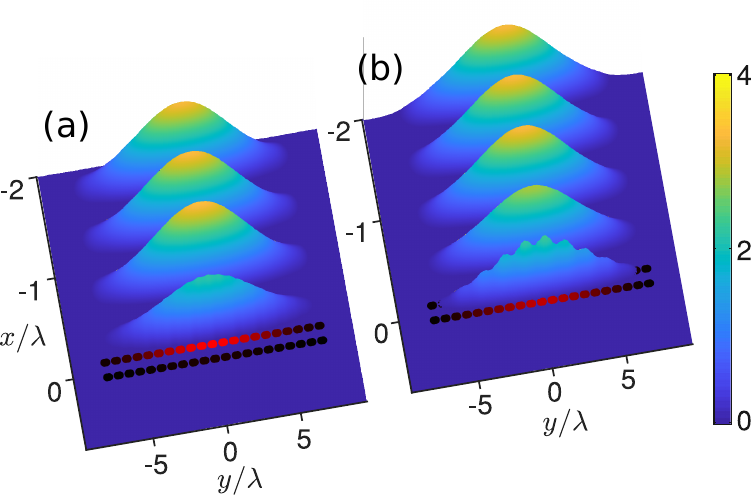}
  \caption{Incident and reflected light from the resonances of (a) electric and (b) magnetic dipole modes, coupled to a cavity photon to create a phase-cat superposition state of a reflected photon.
(a) Resulting standing wave for $\tilde{\delta}=0$, when no cavity photon is present on resonance with the electric dipole mode at $\delta_1\approx 4\gamma$ ($\upsilon_1\approx 1.4\gamma$) and (b) $\tilde{\delta}=-3.6\gamma$, shifting the resonance to the magnetic-dipole mode at $\delta_2 \approx 7.6\gamma$ ($\upsilon_2\approx 0.16\gamma$). The position of the array is illustrated by a single line of the bilayer, with polarization amplitude $\sum_\mu|\Pc_\mu^{(j)}|^2$ ranging from 0 (black) to maximum (red).  $21\times 21 \times 4$ lattice with  $a_y=a_x=0.15\lambda$, $d_y=d_z=5a_y$, and Gaussian illumination with beam waist $w_0=6.4\lambda$. To smooth divergent near-field contributions a Gaussian convolution of width $\lambda/2$ is applied to all electric field data. }
  \label{fig:dmmm}
\end{figure}

Here we show how the ac Stark shift of the cavity field can be used to shift the resonances of the uniform electric dipole mode and the synthesized magnetic dipole mode.
In this way, the system can act as a quantum-optical switch that changes the phase of the reflected photon. In a superposition state of the cavity photon, the coupling then creates a phase-cat state for the reflected photon.
Total reflection from an array of electric dipoles occurs with a well-known $\pi$ phase shift~\cite{Tretyakov,Laroche2006,Abajo07,CAIT}, equivalent to reflection from a perfect electrical conductor~\cite{Tretyakov}. 
This has been verified in experimental realizations in nanophotonics with reflection close to $100\%$~\cite{Moitra2014,Moitra2015}, and has similarly been highlighted for atoms~\cite{Bettles2016,Facchinetti16,Shahmoon} and observed in an optical lattice~\cite{Rui2020}.
Due to the anti-symmetry of the $y$ component of the magnetic dipole or electric quadrupole mode, the back-scattered field amplitude is the negative of the forward scattered one, and so this phase shift is absent~\cite{Ginn12,Schwanecke06,Liu14,Alaee18,Sievenpiper99,Lin16}.
We consider an array coupled to a cavity standing wave with $\vec{k}_s=k_s\unitvec{x}$ for arbitrary $k_s$ and $\phi=0$, as shown in Fig.~\ref{fig:geo}(b), or alternatively $\vec{k}_s=(\pi/d_z)\unitvec{z}$, and polarization $\unitvec{e}_y$. In either case the linear polarization creates an equal shift of the $m=\pm 1$ levels on all atoms (the $m=0$ state does not play a role for the $y$-polarized incident light). This results in an incident forward-propagating state $\ket{\zeta,\unitvec{x}}$ being reflected into a coherent state whose phase is entangled with the collective mode occupation, and hence the cavity photon, with a respective amplitude $\pm\zeta$, i.e.\ 
\begin{equation}
\ket{\zeta,\unitvec{x}}\rightarrow c_0\ket{0}\ket{-\zeta,-\unitvec{x}} + c_1 \ket{1}\ket{\zeta,-\unitvec{x}}.
\end{equation}
The resulting standing wave, formed from interference between the incident and reflected field, is shown in Fig.~\ref{fig:dmmm}. In (a), the cavity is empty and the electric dipole mode is strongly excited, resulting in a $\pi$ phase shift in reflection and a node in the standing wave at $x=0$. In (b), the presence of a cavity photon shifts the magnetic dipole mode on resonance at the same laser frequency, resulting in zero phase shift in the reflected light and an antinode at $x=0$. The reflected intensities in Fig.~\ref{fig:dmmm} are calculated for steady-state illumination, but for sufficiently slowly varying single-photon pulses, with $\gamma\tau\gtrsim 10$, the images become indistinguishable. The efficiency of the reflection can be increased by increasing the lattice size, resulting in more collimated scattering and a better spatial overlap with the Gaussian incident field.

\subsection{Huygens' surface entanglement}
\label{sec:Huygens}

An atomic Huygens' surface, allowing for arbitrary wavefront manipulation of transmitted light, can be engineered by a non-uniform variation of the level shifts of different atoms of the bilayer array. 
Here we propose an atomic Huygens' surface that is considerably simpler and more experimentally feasible than previous proposals~\cite{Ballantine20Huygens}, being only based on the level shifts generated by a single standing wave and selective filling of a fully regular, rectangular, bilayer lattice, as described in Sec.~\ref{sec:magnetism}. Specifically, we show how to achieve quantum control of the transmission by uniform level shifts, providing a quantum-optical switch, able to manipulate, sort, and redirect light.

Huygens' surfaces are based on the Huygens' principle: each point in a propagating wave acts as an independent source of spherical wave-fronts, which interfere to construct the observed field~\cite{Huygens}. 
A more rigorous formulation, which models each point as a pair of crossed electric and magnetic dipoles, explains why these sources only contribute to the forward propagating wave~\cite{Love1901,Schelkunoff36}, as scattered light from each source interferes destructively in the backward direction, suppressing the reflection. The constructive interference between the dipoles in the forward direction can then achieve full $2\pi$ phase control of the transmitted light.
The magnetic dipole can also be replaced by a multipole of similar anti-symmetry (electric quadrupole, magnetic octopole, etc.), with back-scattered light amplitude being the negative of the forward scattered one, allowing for the desired interference with the electric dipole radiation~\cite{Liu17mm}. 

Numerically, we find many different solutions which generate an atomic Huygens' surface.
For the simplified geometry considered here, optimal solutions consist of simultaneous excitation of the electric dipole mode and the uniform electric quadrupole mode, shown in Fig.~\ref{fig:modes}(c). Interpreting the physics in terms of the two-mode model of Eq.~(\ref{eq:tmm}), we now take the uniform electric quadrupole mode to be $\Pc_2$, while $\Pc_1$ denotes the uniform electric dipole mode as before. A variation in the relative shifts of the $m=\pm 1$ levels between different atoms in each unit cell is then required to couple $\Pc_1$, for which all the dipoles in a unit cell are parallel, to $\Pc_2$, for which this is not the case (Appendix~\ref{sec:tmm}). The level shifts lead to non-zero $\bar{\delta}$ in Eq.~(\ref{eq:tmm}) and allow both modes to be excited at the same resonance frequency. To maintain experimental feasibility, we keep these shifts identical in the $x$ direction but varying between the atoms at relative positions $y=\pm a_y/2$, and take the quickly varying level shifts across a unit cell to be controlled by a constant classical laser field. This leaves three independent $\Delta_{\pm}^{(j)}$, up to an overall shift in the resonance frequency, which can be controlled by a standing wave with $\vec{k}_s = (\pi/d_y)\unitvec{y}-(\pi/d_z)\unitvec{z}$, illustrated in Fig.~\ref{fig:geo}(c), as the phase $\phi$ directly controls the relative intensity across the cell. While the level shifts vary only in the $y$ direction, this choice of $\vec{k}_s$ allows for both linear and circular polarization components in the $xy$ plane. Then the intensity, phase, and polarization of the standing wave, along with a uniform linear Zeeman splitting from a background magnetic field, provide more than the necessary three degrees of freedom.

\begin{figure}[htbp]
  \centering
   \includegraphics[width=\columnwidth]{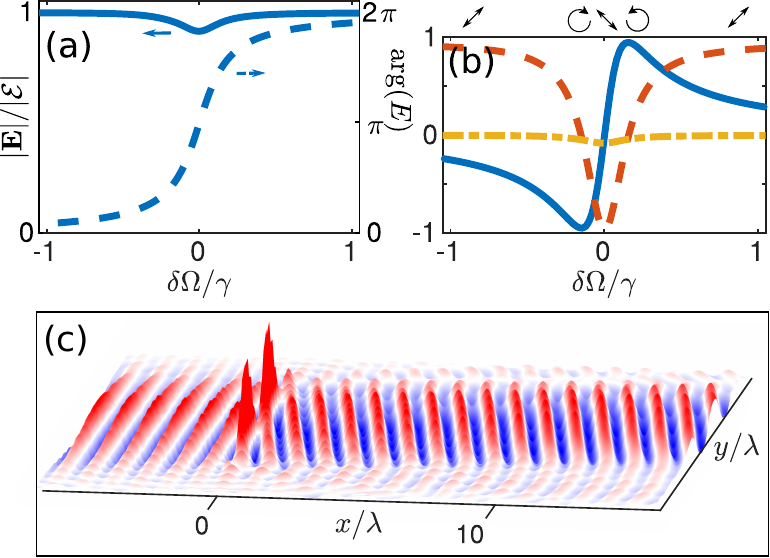}
   \vspace{-0.6cm}
  \caption{ Operation and application of a quantum-controlled Huygens' surface. (a) Magnitude and phase of transmitted light as a function of overall detuning $\Delta\Omega$. (b) Unnormalized Stokes' parameters $V$ (solid), $U$ (dashed), and $Q$ (dot-dashed) of transmitted light for Huygens' surface. As the phase of the $y$ component varies from $0$ to $2\pi$ relative to the $z$ component, the polarization state also varies, as illustrated above the figure.  (c) Beam steering using Huygens' surface controlled cavity photon. $20\times20\times 4$ lattice with parameters for (a,b) given by $a_y=a_x=0.11\lambda$, $d_z=0.8\lambda$, $d_y=7a_y$, and incident plane wave $\mathcal{E}=1$, and for (c) given by $a_y=0.11\lambda$, $a_x=0.22\lambda$, $d_z=0.75\lambda$, $d_y=7a_y$ and input Gaussian beam with waist radius $6.4\lambda$. $\Delta_{+}^{(j)}=-2\gamma$, $\Delta_{-}^{(j)}=-100\gamma$ ($j=1,2$); $\Delta_{+}^{(j)}=-88\gamma$, $\Delta_{-}^{(j)}=-4\gamma$ ($j=3,4$), with atomic labels illustrated in Fig.~\ref{fig:modes}.  }
  \label{fig:polhs}
  \end{figure}

Numerically calculated transmission is shown in Fig.~\ref{fig:polhs}(a), demonstrating a full $2\pi$ phase range. The minimum transmission is $84\%$ at the center of the resonance but can be optimized close to 100\% by allowing the level shifts to vary in the $x$ direction or small shifts in the atomic positions.  While the incident light couples to both modes, it primarily drives the uniform $\Pc_1$. Then, for the solution presented in Fig.~\ref{fig:polhs}, the alternating sign of the effective Zeeman splitting $\Delta_{+}^{(j)}-\Delta_{-}^{(j)}$ between the pair of atoms labeled $(1,2)$ in Fig.~\ref{fig:modes}, and the pair labeled $(3,4)$, couples the uniform $y$ polarization to the varying $x$ polarization of the quadrupole mode, producing an effective $\bar{\delta}$ in Eq.~(\ref{eq:tmm}) and allowing both modes to be excited at a single resonance frequency. 

Remarkably, an atomic Huygens' surface allows quantum control by a simple uniform level shift applied to all the atoms, which could be achieved by an interacting qubit, such as a resonance shift generated by a Rydberg atom~\cite{Zhang21,Cardoner21,Bekenstein2020}.
Here we consider a cavity photon as in Secs.~\ref{sec:storage} and ~\ref{sec:cat}.
While the classical field enables the operation of the Huygens' surface, the additional quantized cavity photon with $\vec{k}_s=k_s\unitvec{x}$, identical to that considered in Sec.~\ref{sec:cat}, adds an overall shift $\tilde{\delta}$, moving the resonance shown in Fig.~\ref{fig:polhs}(a) and so allowing switching between two different phases of transmission.  

The overall phase control allows the Huygens' surface to be used as an effective variable waveplate by delaying the incoming $y$ polarization, which couples to the Huygens' surface resonance, relative to the $z$ polarization, which does not. The resulting Stokes polarization parameters are shown in Fig.~\ref{fig:polhs}(b) as a function of the detuning $\Delta\Omega$, or equivalently for varying $\tilde{\delta}$ at fixed $\Delta\Omega$. Here, positive $V$, $U$, and $Q$ correspond to right-circular, diagonal, and horizontal polarization, respectively, and negative values to the respective orthogonal polarization. The waveplate has full $2\pi$ control of the relative phase of fast and slow axes, allowing for the transmitted polarization to vary over a full great-circle on the Poincar\'{e} sphere. Then, for diagonal polarized input $\unitvec{e}_{\rm in}=\unitvec{e}_y+\unitvec{e}_z$, the ac Stark shift of the cavity field can switch between two different phase delays, or equivalently two orthogonal polarizations of the transmitted photon. A shift $\tilde{\delta}=0.3\gamma$, for example, leads to entanglement
$\ket{\zeta,d}\rightarrow c_0\ket{0}\ket{\zeta,-}+c_1\ket{1}\ket{\zeta,+}
$
where an incoming state, with amplitude $\vecg{\zeta}(t)$ and diagonal polarization, is converted to an entangled state of right or left circular polarization.

More complex wavefront engineering can be achieved if the level shifts also have some slowly varying spatial dependence. This can be done by using a magnetic field gradient or a classical field, which could be produced by a spatial light modulator, to impart the spatial structure. The operation of the Huygens' surface can still be controlled by the cavity photon uniformly shifting the resonance. When the resonance is at $\Delta\Omega=0$ level shifts in a range between $-\gamma$ and $\gamma$ correspond to a full $2\pi$ phase variation. When the resonance is shifted to $|\Delta\Omega|\gtrsim 2\gamma$, however, the same range of level shifts results in an approximately constant zero phase, allowing the photon to turn on or off the wavefront shaping. A linear phase gradient leads to a deflection in the direction of the transmitted beam~\cite{Pfeiffer13,Liu17bs,Ni12}, allowing the light to be sorted into different directions for further processing, all dependent on the presence of a cavity photon which can itself depend on the result of prior quantum calculations. Figure~\ref{fig:polhs}(c) shows beam steering by the atomic array of $10^\circ$. Other spatially varying phase profiles could, in principle, also be used to control beam focusing, or the generation of orbital angular momentum~\cite{Ballantine20Huygens}, leading to similar entanglement of outgoing coherent states transmitted into different modes
$
\ket{\zeta}\rightarrow c_0\ket{0}\ket{\zeta,0}+c_1\ket{1}\ket{\zeta,1}
$,
where $\ket{\zeta,n}$ can refer to two general outgoing modes labeled by $n=0,1$.

\section{Concluding remarks}

We have shown how cooperative responses in planar arrays of atoms can be harnessed for achieving optical manipulation that is reminiscent of artificially fabricated metasurfaces, but with single-photon quantum control based on atomic physics technology.  
This allows for engineering controllable many-atom quantum systems with light-induced couplings between spatially delocalized collective excitation eigenmodes which result from the scattered light mediating strong interactions between the atoms. Despite the complexity of the system, these collective modes, however, can be coupled to qubits in ways that mathematically are similar to single-particle coupling.
Due to their exceptional properties of effectively 1D collimated light propagation,
strong coupling to light with a large optical cross-section, and the ability to manipulate the atomic states, planar atomic arrays provide a promising light-matter interface for quantum information processing platforms.   
To illustrate this, we have shown how a time-dependent single-photon pulse can be deterministically stored in, and retrieved from, a subradiant mode of an atomic array, with near-unit efficiency. Extreme wavefront control of a photon is achieved with an atomic
 Huygens' surface. A quantized cavity field is then used to control the phase, polarization, and direction of propagating photons. 
The proposal allows for planar atomic arrays to form nodes as part of a larger quantum-information architecture, transmitting the quantum state of the cavity to the outgoing light. As conventional metasurfaces have had extraordinary success in controlling the classical properties of light, the coherent quantum control of light by atomic arrays opens new avenues for controlling and manipulating light at the single-photon level.    

 \begin{acknowledgments}
We acknowledge financial support from the UK EPSRC (Grant Nos. EP/S002952/1, EP/P026133/1).
\end{acknowledgments}

\appendix

\section{ac Stark shift}
\label{sec:Stark}

We assume an ac Stark shift due to either a classical coherent field or a cavity photon, following from coupling between the $J^\prime=1$, $m=\pm 1,0$ states, $\ket{e_{\mu j}}$ with frequency $\omega_{\mu}^{(j)}$, and other off-resonant states which we label $\ket{k}$, with frequency $\omega_k$. The emergence of intensity-dependent level shifts can be illustrated by
adiabatically eliminating the additional states in the Jaynes-Cummings Hamiltonian consisting of the control field $\hat{H}_s$, atom $\hat{H}_a$, and light-atom  coupling $\hat{H}_{\rm int}$ parts,
\begin{align*}
\hat{H}_s &= \omega_s \hat{a}_s^\dagger \hat{a}_s,\quad \hat{H}_{\rm int} = \sum_{\mu j,k} g_{\mu k}^{(j)} \ket{k}\bra{e_{\mu j}}\hat{a}_s + {\rm H.c.},\\
\hat{H}_a &= \sum_k \omega_{k} \ket{k}\bra{k} + \sum_{\mu j} \omega_{\mu}^{(j)} \ket{e_{\mu j}}\bra{e_{\mu j}}, 
\end{align*}
where $g_{\mu k}^{(j)}$ is the light-induced coupling strength of the transition from $\ket{e_{\mu j}}$ to $\ket{k}$. For large detuning $\big|\omega_s-\big(\omega_{k}-\omega_{\mu }^{(j)}\big)\big|\gg g_{\mu k}^{(j)}$ the off-resonant states $\ket{k}$ are adiabatically eliminated.
This leaves the effective interaction contribution  $ \delta_{\mu}^{(j)}  \hat{a}_s^\dagger \hat{a}_s  \ket{e_{\mu j}}\bra{e_{\mu j}}$
where $\delta_{\mu }^{(j)} =\sum_k \big(g_{\mu k}^{(j)}\big)^2/(\omega_{k}-\omega_s)$. 
For the classical control field, we can replace $\hat{a}_s^\dagger \hat{a}_s$ by the classical intensity contribution, and the larger level shifts required in Sec.~\ref{sec:Huygens} can be achieved with high-intensity lasers.
 
While the general form of the level shifts can be derived from the Jaynes-Cummings Hamiltonian, the precise values of the transition strengths $g_{\mu k}^{(j)}$ depend on the selection rules and intensity. A more rigorous second order perturbation calculation shows that for the classical standing wave described in Sec.~\ref{sec:ints}, the general form of the ac Stark shift can be decomposed into different components as~\cite{Schmidt16,Rosenbusch09,LeKien13} 
\begin{align}
\delta\omega_{\mu }^{(j)} &= -\frac{1}{4}|\mathcal{E}_s|^2\cos^2{(\vec{k}_s\cdot\vec{r}_j+\phi)} \left[\vphantom{\frac{(3\mu^2-1)}{2}}\alpha^s(\omega)\right. \nonumber \\ &\phantom{==}+ \left. C \frac{\mu}{2} \alpha^v(\omega)    -D\frac{(3\mu^2-2)}{2}\alpha^T(\omega)\right],
\end{align} 
where $C=|u_{-}|^2-|u_{+}|^2$ parameterizes the degree of circular polarization of $\unitvec{e}_s=\sum_\mu u_\mu \unitvec{e}_\mu$, $D=1-3|u_0|^2$, and $\alpha^{\rm s}(\omega)$, $\alpha^{\rm v}(\omega)$, and $\alpha^{\rm T}(\omega)$ are the scalar, vector, and tensor polarizabilites, respectively. Thus, light with a component of circular polarization in the $xy$ plane leads to a $\mu$ dependent splitting $\Delta_{+}^{(j)}\neq\Delta_{-}^{(j)}$, while a linear polarization component leads to an equal shift of the $m=\pm 1$ levels in the same direction. The different contributions are captured in the function $U$ in Sec.~\ref{sec:ints}.

\section{Two-mode model}
\label{sec:tmm}

The role of the level shifts on the mode coupling in the two-mode model [Eq.~(\ref{eq:tmm})] can be illustrated for a single atom  with polarization amplitudes $\Pc_\mu$. In the basis of circular polarization, the amplitudes evolve independently as
\begin{equation}
\partial_t\Pc_{\mu} = (i\Delta_{\mu}+i\Delta\Omega-\gamma)\Pc_{\mu}  + \zeta_\mu(t),  
\end{equation}
with $\zeta_\mu(t)=\unitvec{e}_\mu^\ast\cdot\unitvec{e}_{\rm in} f(\vec{r})\exp{(-t^2/\tau^2)}$. We take $\unitvec{e}_{\rm in}=\unitvec{e}_y$ and  rewrite these equations in terms of the Cartesian components
to find
\begin{align*}
\partial_t \Pc_x &= (i\tilde{\delta}+i\Delta\Omega-\gamma)\Pc_x +\bar{\delta}\Pc_y, \\
\partial_t \Pc_y &= (i\tilde{\delta}+i\Delta\Omega-\gamma)\Pc_y -\bar{\delta}\Pc_x + \zeta_y(t), 
\end{align*}
with  $\tilde{\delta}=(\Delta_{+}+\Delta_{-})/2$, and $\bar{\delta}=(\Delta_{+}-\Delta_{-})/2$. The incident light drives $\Pc_{y}$, but the $m=\pm 1$ level shifts couple $\Pc_y$ to $\Pc_x$. The $J=0\rightarrow J^\prime=1$ transition is isotropic in the absence of level shifts and any orthogonal basis is equivalent. This isotropy is broken by the level shifts and $\Pc_{x,y}$ are no longer eigenstates, causing the coupling between these modes~\cite{Facchinetti16}. 

For a single layer lattice, an incident light with a uniform phase profile drives almost uniquely the uniform mode with all atoms oscillating in-phase along the $y$ direction. As in a single-atom case, the same level shifts in each atom then couple $\Pc_y$ and $\Pc_x$, leading to an overall coupling between the uniform (in terms of the phase profile) in-plane mode and the uniform out-of-plane mode. The overall response of the array to light has been shown to be well-described by the two-mode model Eq.~(\ref{eq:tmm})~\cite{Facchinetti16,Facchinetti18,Ballantine20ant}
with $\Pc_{1,2}=\Pc_{y,x}$  and $\zeta_2=0$.

For the atomic bilayer, the coupling is more complicated, and the required level shifts must be found numerically, but we still find a good qualitative agreement with Eq.~(\ref{eq:tmm}) with $\Pc_1$ the uniform electric dipole mode and $\Pc_2$ being one of either the uniform magnetic dipole mode or the uniform electric quadrupole mode. The parameter $\tilde{\delta}$ is defined by the sum over the level shifts in  a unit cell, $\tilde{\delta}=\sum_{\mu=\pm1}\sum_{j=1}^{4} \Delta_{\mu}^{(j)}/8$. The coupling $\bar{\delta}$ depends on the particular solution and whether the magnetic dipole or electric quadrupole mode is targeted, but requires two key features; a difference $\Delta_{+}^{(j)}\neq\Delta_{-}^{(j)}$ to couple the $y$ and $x$ polarizations on each atom, and a difference $\Delta_{\mu}^{(j)}\neq\Delta_{\mu}^{(k)}$ for $j\neq k$ between level shifts on different atoms within each unit cell to couple the electric dipole mode, where the dipoles are parallel, to the magnetic dipole or electric quadrupole mode, where the orientations of the dipoles between the atoms differ. For the case considered in Sec.~\ref{sec:Huygens} where the level shifts vary only in the $y$ direction, the difference in level splitting $\bar{\delta}= (\bar{\delta}^{(1)}-\bar{\delta}^{(3)})/(2\sqrt{2})$ where $\bar{\delta}^{(j)}=(\Delta_{+}^{(j)}-\Delta_{-}^{(j)})/2$ with the atoms labeled as in Fig.~\ref{fig:modes}.

\section{Quantum and classical entanglement}
\label{sec:entanglement}

We analyze the collective response of the arrays to a single photon and therefore restrict the quantum dynamics only to the single excitation sector of the Hilbert space.
The single excitation sector, described in Sec.~\ref{sec:ints}, is represented by the entangled state~\cite{Ballantine20ant} 
\begin{align}
\ket{\Psi} &=\sum_{\mu j} \Pc_{\mu}^{(j)}\hat{\sigma}_{\mu j}^{+}\ket{G}\nonumber\\
&= \sum_{\mu j} \Pc_{\mu}^{(j)} \ket{g_1,\ldots,g_{j-1},e_{j,\mu},g_{j+1},\ldots,g_N}.
\label{eq:quantum}
\end{align}
The dynamical equations for this state (Sec.~\ref{sec:ints}), namely $\dot{\vec{b}}=i(\mathcal{H}+\mathcal{H}^{\prime})\vec{b}+ \vecg{\zeta}(t)$, can also describe the classical response of the array to a coherent probe in the limit of low light intensity (LLI). Here, we discuss the difference between these two descriptions, and show that only in the quantum case do we develop entanglement between the array and the cavity. 

In the limit of LLI the equations for atoms and light are derived to first order in light field amplitude by keeping the terms that include at most one of the amplitudes for the incident light or atomic polarization~\cite{Ruostekoski1997a}. The resulting dynamics
constitutes a coupled-dipole model for a collection of coupled linear harmonic oscillators (radiatively coupled dipoles) in an incident field, providing an exact solution for coherently driven stationary atoms in the limit of LLI for any given atom statistics~\cite{Javanainen1999a,Lee16}. 

The equations for the LLI coupled-dipole model are classical, and the wavefunction for the state factorizes into a product of separable contributions from different atoms
\begin{equation}
\ket{\Psi} = \prod_j \left( c_g^{(j)}\ket{g_j}+\sum_\mu c_{e,\mu}^{(j)}\ket{e_{j,\mu}}\right).
\label{eq:classical}
\end{equation}
The limit of LLI (classical coupled-dipole model) then corresponds to keeping only terms to first order in $c_{e,\mu}^{(j)}$. This yields $|c_{e,\mu}^{(j)}|^2=0$ and $|c_g^{(j)}|^2=1$ when analyzing Eq.~\eqref{eq:classical}.
For simplicity of the notation, in the following discussion we drop the polarization label $\mu$ that refers to the different hyperfine levels of the electronic excitation of the atoms. The classical state then becomes a simple product state
of the amplitudes $( c_g^{(j)}\ket{g_j}+ c_{e}^{(j)}\ket{e_{j}})$.
These amplitudes evolve with the same dynamics as the quantum amplitudes $\Pc^{(j)}$, with $b_j=  [c_g^{(j)}]^\ast c_{e}^{(j)}$. 
In this description, the amplitudes correspond to off-diagonal elements in the single-particle density matrix
\begin{equation}
\hat{\rho}_a^{(j)} = \begin{pmatrix}
\rho_{ee} && \rho_{eg} \\ \rho_{ge} && \rho_{gg}
\end{pmatrix}=\begin{pmatrix}
0 &&   [c_g^{(j)}]^\ast c_{e}^{(j)} \\   c_g^{(j)} [c_{e}^{(j)}]^\ast && 1
\end{pmatrix},
\end{equation}
i.e.\ $\bra{e_j}\hat{\rho}_{a}^{(j)}\ket{e_j}=0$ and $\bra{g_j}\hat{\rho}_{a}^{(j)}\ket{g_j}=1$. Then the collective atom state,
\begin{equation}\label{eq:classrho}
\hat{\rho}_a = \hat{\rho}_a^{(1)}\otimes \hat{\rho}_a^{(2)}\otimes \ldots \otimes \hat{\rho}_a^{(N)},
\end{equation}
analogously to Eq.~\eqref{eq:classical},
factorizes into separable contributions from different atoms, with no quantum entanglement. Hence, the probability to find the array in the excited state is zero, and the amplitudes $ [c_g^{(j)}]^\ast c_e^{(j)}$ give the coherence only between the excited state $\ket{e_j}$ and the collective ground state, in contrast to the quantum state of Eq.~\eqref{eq:quantum} for which the probability of the lattice to contain a single excitation is $\sum_j|\Pc^{(j)}|^2 = 1$. When we include the finite probability $p_G$ of the
excitation to decay to the ground state, we can write the quantum density matrix (Sec.~\ref{sec:ints}) as $\hat{\rho}_a=\ket{\Psi}\bra{\Psi}+p_G\ket{G}\bra{G}$.

The similarity of the dynamical equations of motion for the quantum and classical cases can be considered surprising taking into account the dramatically different representations of the physical states in Eqs.~\eqref{eq:quantum} and~\eqref{eq:classical}.
Such differences are well recognized in quantum optics textbooks~\cite{CarmichaelVol2} and collective scattering literature~\cite{SVI10}, and do not mitigate the quantum properties of the entangled Eq.~\eqref{eq:quantum}. The entanglement manifests itself in particular in the coupling of
the array excitation to the cavity photon, as discussed in the examples of Secs.~\ref{sec:storage},~\ref{sec:cat}, and ~\ref{sec:Huygens}. We show how the quantum treatment of a single photon excitation exhibits quantum entanglement, while the classical coupled-dipole model does not.
We write  the pure quantum state of a single-excitation atomic array and a cavity as
$\hat{\rho} = \sum_{nm}c_n c_m^\ast \ket{\Psi,n}\ket{n}\bra{\Psi,m}\bra{m}$, where $\ket{\Psi,n}=\sum_j\mathcal{P}^{(j,n)}\sigma_j^{+}\ket{G}$, and $\ket{n}$ is the cavity state. The amplitudes $\mathcal{P}^{(j,n)}$ are solved separately for $n=0$, corresponding to no cavity photon, and $n=1$, when the photon, and the resulting level shifts are present. As in Sec.~\ref{sec:storage}, we consider a single-layer with coupling between the uniform in-plane mode $\Pc_1$ and the uniform out-of-plane mode $\Pc_2$.   
Then $n=0$ results in $\bar{\delta}=0$ in Eq.~(\ref{eq:tmm}), and no coupling between the two modes, while $n=1$ corresponds to $\bar{\delta}\neq 0$, allowing the dipoles to rotate and coupling $\Pc_1$ to $\Pc_2$. The full dynamics are calculated  separately for both cases and the two sets of amplitudes used to reconstruct the full density matrix. The classical state is similarly described by $\bra{e_j}\bra{n}\hat{\rho}\ket{G}\ket{m} =  [c_g^{(j,n)}]^\ast  c_{e}^{(j,n)}$, again calculated separately for $n=0,1$.

\begin{figure}[htbp]
  \centering
   \includegraphics[width=0.7\columnwidth]{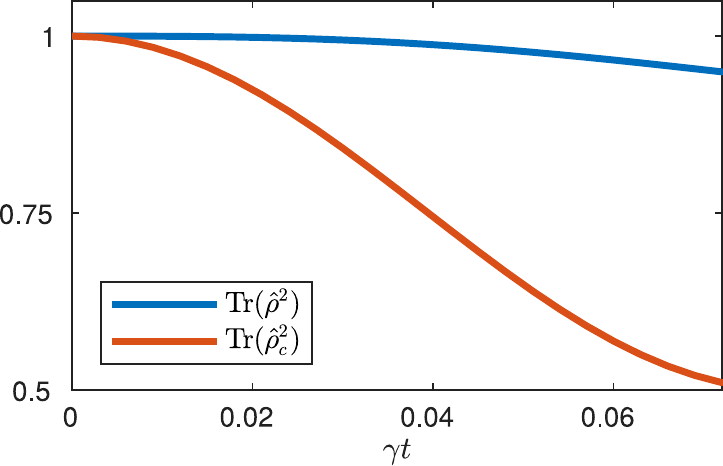}
  \caption{Entanglement measure calculated numerically for pure quantum single-photon excitation of $27\times 27$ lattice, with $\bar{\delta}=20\gamma$, and cavity state $(\ket{0}+\ket{1})/\sqrt{2}$. }
  \label{fig:entanglement}
\end{figure}

We can calculate the bipartite entanglement~\cite{Palmer05} that provides the conditions for separability of the density matrix by comparing the purity of the whole density matrix $\hat{\rho}$ with either the reduced density matrix $\hat{\rho}_c={\rm Tr}_a(\hat{\rho})=\sum_j\bra{e_j}\hat{\rho}\ket{e_j}+\bra{G}\hat{\rho}\ket{G}$ of the cavity subsystem when the partial trace is taken over the atomic degrees of freedom, or the reduced density matrix $\hat{\rho}_a={\rm Tr}_c(\hat{\rho})=\sum_n \bra{n}\hat{\rho}\ket{n}$ of the atoms with the partial trace taken over the cavity. If there is no entanglement and the density matrix can be factorized into atomic and cavity subsystems, i.e.\ $\hat{\rho}=\hat{\rho}_a\otimes\hat{\rho}_c$, then ${\rm Tr}(\hat{\rho}^2)\leq {\rm Tr}(\hat{\rho}_c^2)$~\cite{Palmer05,Alves04}.

We take the initial pure state at $t=0$ to consist of exactly one excitation in the out-of-plane mode $\Pc_2$, as in Fig.~\ref{fig:decay}.
We show here the case of large level shifts $\bar{\delta}=20\gamma$, such that coupling between the two modes occurs on a time scale short compared to the decay rate of either mode, and the state remains approximately pure.  
The result is shown in Fig.~\ref{fig:entanglement} for the density matrix calculated numerically for a large lattice, both for the full system and for the cavity subsystem. As the evolution of the atoms depends on the state of the cavity, the density matrix quickly shows entanglement between the atoms and the cavity, with ${\rm Tr}(\hat{\rho}^2) > {\rm Tr}(\hat{\rho}_c^2)$. This result is explained by the two-mode model, Eq.~(\ref{eq:tmm}). Solving this equation in the same limit where decay is neglected and the state remains pure, and reconstructing the density matrix gives
\begin{equation}
{\rm Tr}(\hat{\rho}^2) \simeq1, \quad
{\rm Tr}(\hat{\rho}_c^2) \simeq \half \left[1+\cos^2(\bar{\delta} t)\right],
\end{equation}
which closely matches the numerical results in Fig.~\ref{fig:entanglement} for short time scales.
Taking the classical state and repeating the calculation gives 
\begin{equation}
{\rm Tr}(\hat{\rho}^2) \simeq 1+\mathcal{O}(|c_e|^2), \quad
{\rm Tr}(\hat{\rho}_c^2) \simeq 1,
\end{equation}
which, since the requirement for the classical limit of LLI is to keep the amplitude terms $c_e$ only to first order, results in ${\rm Tr}(\hat{\rho}^2) = {\rm Tr}(\hat{\rho}_c^2)=1$ corresponding to a pure, separable state.

%

\end{document}